\documentclass[floats,twocolumn]{revtex4}
\begin{document}

\title{Reply to Comment on ``Signatures of surface states in bismuth at high magnetic fields''}

\author{Babak Seradjeh}
\affiliation{Department of Physics, University of Illinois, 1110 West Green St, Urbana 61801, USA.}
\author{Jiansheng Wu}
\affiliation{Department of Physics, University of Illinois, 1110 West Green St, Urbana 61801, USA.}
\author{Philip Phillips}
\affiliation{Department of Physics, University of Illinois, 1110 West Green St, Urbana 61801, USA.}

\maketitle

In a Comment~\cite{Beh09a}, Behnia contends that the surface theory put forward in our recent Letter~\cite{SerWuPhi09a} as an alternative explanation of the anomalous peaks observed in Nernst measurement on a single-crystal bismuth sample at high ($\agt9$ T) magnetic fields~\cite{BehBalKop07a} is not consistent with the order of magnitude and shape of the anomalous peaks observed in the experiment. We explain in this Reply why this contention is not true.

The value of $\ell\sim1~\mu$m that was reported in our Letter~\cite{SerWuPhi09a} (denoted by $e$ in~\cite{Beh09a}) is not the ``assumed'' extension of the surface states into the bulk nor is it ``implausible'' as Behnia writes; rather, it is a relevant length scale of the problem in the presence of surface states: $\ell=n'/n$ where $n$ and $n'$ are the density of electrons in the bulk and at the surface, respectively. Behnia also claims in passing that we mistook the width of the sample for its thickness (differing by a factor $\sim 2$), while Fig.~2(b) of Ref.~\onlinecite{BehMeaKop07a} clearly indicated otherwise (compare with Fig.~1 of Ref.~\onlinecite{Beh09a}). This slight difference does not have any physical consequence for the purpose at hand.

In a metal the Nernst signal can be written as 
\begin{equation}
S_{xy}=\frac{\pi^2}3\frac{k_B^2 T}{e}\left.\frac{\partial\theta_H}{\partial\epsilon}\right|_{\epsilon=\epsilon_F},
\end{equation}
where $k_B$ is the Boltzmann constant, $e$ the elementary charge, $T$ the temperature, $\epsilon_F$ the Fermi energy and $\theta_H=\tan^{-1}(\sigma_{xy}/\sigma_{xx})$ is the Hall angle. It is then easy to see that when $\theta_H\approx 0$ or $\pi/2$ (as is usually the case) the leading correction to the bulk Nernst signal due to the surface states is
\begin{equation}
\left|\frac{\delta S_{xy}}{S_{xy}}\right| \approx \frac{\ell}w \approx 10^{-3},
\end{equation}
where $w$ is the thickness of the sample along the direction of the magnetic field. However, this estimate only applies to the background signal, not the amplitude of the quantum oscillations. For instance, a numerical calculation for an electron gas in two dimensions~\cite{NakHatShi05a} found peaks in the Nernst signal due to the edge states as large as 0.07 mV/K for lower Landau levels, while the background signal, strictly speaking, vanishes as $T\to0$. So, we do not agree that there is an inconsistency between our theory and the bismuth data of Ref.~\onlinecite{BehBalKop07a} where the background signal is $\sim 1$ mV/K and the strength of the anomalous peaks is $\sim 0.1$ mV/K. We note also that the peaks caused by bulk Landau levels for fields close to but below $9$ T are stronger than the anomalous peaks by an order of magnitude.

Behnia has argued that the shape of the peaks in the Nernst signal can be used to determine whether the signal arises from the bulk or the surface. However, without a proper numerical analysis it is dangerous to rely on the shape of the peaks as a measure of dimensionality in a system with mixed bulk and surface states. The shape could be useful only if the background bulk signal is carefully subtracted. The shape of the anomalous peaks in the data presented in Ref.~\onlinecite{BehBalKop07a} could support both a surface and a bulk scenario depending on how the background is subtracted. We point out that the locations of the peaks obtained in our Letter~\cite{SerWuPhi09a} do in fact seem to be centered in the middle of a peak and a trough, a quality that Behnia argues is a feature of quantum oscillations of the Nernst signal in some two-dimensional system.

It is important to note here that a similar estimate for surface corrections to the resistivity gives
\begin{equation}
\left|\frac{\delta\rho}{\rho}\right|\approx \frac \ell w \times \frac{\mu'}{\mu}\alt10^{-5},
\end{equation}
where $\mu$ and $\mu'$ are the mobility in the bulk and at the surface, respectively. This further amplifies why the surface states are much more difficult to detect in a resistivity measurement. Our surface theory nicely explains the featureless and finite profile of the resistivity (see supporting material of Ref.~\onlinecite{BehBalKop07a}) in this range of fields, a fact that poses a significant challenge to any bulk explanation, let alone one that resorts to fractionalization of electrons in an isotropic matter in three dimensions as argued by Behnia {\em et al.}~\cite{BehBalKop07a}.

\end{document}